\documentclass[prb,twocolumn,showpacs,superscriptaddress]{revtex4}

\usepackage{graphicx}
\usepackage{dcolumn}
\usepackage{amsmath}
\usepackage{color}

\newcommand{\CuBiSeOBr}{Cu$_{3}$Bi(SeO$_3$)$_2$O$_2$Br}
\newcommand{\CuBiSeOCl}{Cu$_{3}$Bi(SeO$_3$)$_2$O$_2$Cl}

\begin{document}
\title{THz spectroscopy in the pseudo-Kagome system Cu$_{3}$Bi(SeO$_3$)$_2$O$_2$Br}

\author{Zhe~Wang}
\author{M.~Schmidt}
\affiliation{Experimental Physics V, Center for Electronic
Correlations and Magnetism, Institute of Physics, University of Augsburg, D-86135 Augsburg, Germany}

\author{Y.~Goncharov}
\affiliation{Experimental Physics V, Center for Electronic
Correlations and Magnetism, Institute of Physics, University of Augsburg, D-86135 Augsburg, Germany}
\affiliation{Institute of General Physics, Russian Academy of Sciences, RU-119991 Moscow, Russia}

\author{V.~Tsurkan}
\affiliation{Experimental Physics V, Center for Electronic
Correlations and Magnetism, Institute of Physics, University of Augsburg, D-86135 Augsburg, Germany}
\affiliation{Institute of Applied Physics, Academy of Sciences of Moldova, MD-2028 Chisinau, Republic of Moldova}

\author{H.-A.~Krug von Nidda}
\author{A.~Loidl}
\author{J.~Deisenhofer}
\affiliation{Experimental Physics V, Center for Electronic
Correlations and Magnetism, Institute of Physics, University of Augsburg, D-86135 Augsburg, Germany}

\date{\today}

\begin{abstract}
Terahertz (THz) transmission spectra have been measured as function of temperature and magnetic field on single crystals of \CuBiSeOBr. In the time-domain THz spectra without magnetic field, two resonance absorptions are observed below the magnetic ordering temperature $T_N\sim27.4$~K. The corresponding resonance frequencies increase with decreasing temperature and reach energies of 1.28 and 1.23~meV at 3.5~K. Multi-frequency electron spin resonance transmission spectra as a function of applied magnetic field show the field dependence of four magnetic resonance modes, which can be modeled as a ferromagnetic resonance including demagnetization and anisotropy effects.
\end{abstract}

% 78.40.-q    Absorption and reflection spectra: visible and ultraviolet (for infrared spectra, see 78.30.-j)
% 78.20.-e    Optical properties of bulk materials and thin films (for optical properties related to materials treatment,
%             see 81.40.Tv; for optical materials, see 42.70-a; for optical properties of superconductors,
%             see 74.25.Gs; for optical properties of rocks and minerals, see 91.60.Mk)
% 71.70.-d    Level splitting and interactions (see also 73.20.-r Surface and interface electron states;
% 71.70.Ch    Crystal and ligand fields
% 71.70.Ej    Spin-orbit coupling, Zeeman and Stark splitting, Jahn-Teller effect
% 75.30.Fv    Spin-density waves
% 75.20.Hr    Local moment in compounds and alloys; Kondo effect, valence fluctuations, heavy fermions
% 71.70.Ch    Crystal and ligand fields
% 76.30.-v    Electron paramagnetic resonance and relaxation
% 76.50.+g 	  Ferromagnetic, antiferromagnetic, and ferrimagnetic resonances; spin-wave resonance (see also 75.30.Ds Spin waves)
% 75.10.Hk    Classical spin models
% 78.30.-j 	  Infrared and Raman spectra


\maketitle

\section{Introduction}
Systems composed of S=1/2 spins on a two-dimensional Kagome lattice can be strongly frustrated due to competing exchange interactions, which may lead to interesting ground states and exotic low-energy excitations.\cite{Balents10} Cu$_{3}$Bi(SeO$_3$)$_2$O$_2$\emph{X} (\emph{X} = Cl, Br) is such a system that is composed of Cu$^{2+}$ ions on slightly buckled Kagome layers with weak inter-layer coupling.\cite{Millet01} This system also has an interesting noncollinear magnetic structure as the non-nearest-neighbor exchange paths mediated by the lone-pair ions Bi$^{3+}$ and Se$^{4+}$ could be more efficient than the nearest-neighbor one.\cite{Lemmens03,Pregelj12}

Cu$_{3}$Bi(SeO$_3$)$_2$O$_2$\emph{X} crystallizes in an orthorhombic symmetry with space group \emph{Pmmn}.\cite{Pring90} The unit-cell parameters of \CuBiSeOBr\, are $a=6.390$~{\AA}, $b=9.694$~{\AA}, and $c=7.287$~{\AA}.\cite{Millet01} The structure is characterized by two types of Cu ions, Cu1 and Cu2, with the site symmetries $-1$ and \emph{mm2}, respectively. These Cu ions are surrounded by four oxygen ions forming two types of planar building blocks. Every block shares one corner with each of the two neighboring blocks.\cite{Pregelj12} Neutron-diffraction experiments and magnetic susceptibility measurements have revealed a magnetic ordering transition from a paramagnetic to an antiferromagnetic phase at $T_N\sim$ 27.4~K.\cite{Pregelj12} In the antiferromagnetic state, the spins of Cu ions from neighboring \emph{ab}-layers are aligned antiparallel.\cite{Pregelj12} Within an \emph{ab}-layer, the spins of Cu2 ions are parallel. The spins of Cu1 ions along the \emph{a}-axis are aligned parallel, while those of Cu1 ions along the \emph{b}-axis form a noncollinear order. The inter-layer antiferromagnetic ordering can be switched to ferromagnetic by an applied magnetic field of 0.8~T along the \emph{c}-axis at 2~K.\cite{Pregelj12} This is confirmed by specific-heat measurements in an applied magnetic field.\cite{Guenther12} Strong exchange anisotropy has been found in \CuBiSeOBr\, with the \emph{c}-axis as the easy axis according to the magnetic susceptibility measurements.\cite{Pregelj12} The saturation of magnetization is reached at 0.8~T with the applied magnetic field along \emph{c}-axis, while the saturation field is about 7 and 15~T for the field parallel to \emph{b}- and \emph{a}-axis, respectively. Nevertheless, the nature of the exchange interactions is still unclear, since the additional exchange paths via Cu-O-\emph{Z}-O-Cu~(\emph{Z} = Bi, Se) could be more important than the nearest-neighbor Cu-O-Cu paths.\cite{Deisenhofer06,Das,Wang11b}

In this work we perform temperature-dependent THz time-domain transmission measurements and multi-frequency electron spin resonance transmission experiments on single crystals of \CuBiSeOBr. Four magnetic resonance modes have been resolved from the spectra. In the ferromagnetic phase above 0.86~T, demagnetization and magnetic anisotropy effects have been considered to model the field dependence of the resonance modes in terms of ferromagnetic resonance. This modeling yields a \emph{g}-factor of 2.04(8) and a magnetic anisotropic energy of approximately 0.25~meV.

\section{Experimental details}

Single crystals of \CuBiSeOBr~were grown at
$500-550\,^{\circ}{\rm C}$ by chemical-transport reaction
with bromine as transport agent using polycrystalline material prepared by solid-state reactions from high-purity
binary compounds.\cite{Pregelj12} Time-domain THz transmission measurements were carried out on a $2\times2\times0.3$~mm$^{3}$ single crystal for 3.5~K $<T<$ 300~K using a TPS spectra 3000 spectrometer with $f/2$ focusing optics (TeraView Ltd.). Multi-frequency transmission experiments were performed with an external magnetic field varying up to 7~T in the Faraday configuration (propagating vector $\vec{k}$ $\|$ \textbf{H} $\|$ \emph{c}-axis) on a $3\times3\times1$~mm$^{3}$ single crystal
with backward-wave oscillators covering the frequency range 300 - 490~GHz and a magneto-optical cryostat (Oxford/Spectromag).

\section{Experimental results and discussions}

\subsection{Time-domain THz spectra}
%Figure 1
\begin{figure}[t]
\centering
\includegraphics[width=75mm,clip]{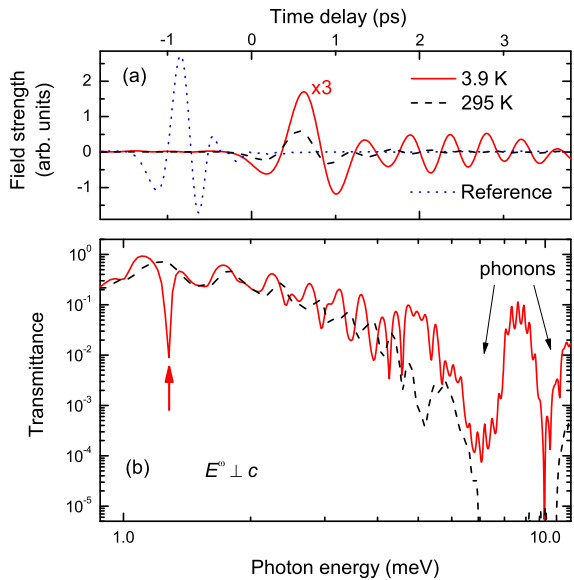}
\vspace{2mm} \caption[]{\label{Fig:TDS} (Color
online) (a) Electric-field transients as a function of time delay measured at 3.9~K and 295~K with $\mathbf{E}^\omega\perp c$. The 3.9~K curve is enlarged by a factor of 3 in order to distinguish from the 295~K curve.  (b) Transmittance as a function of photon energy at 3.9~K and 295~K.}
\end{figure}

Figure~\ref{Fig:TDS}(a) shows the transient electric field through the single crystal of \CuBiSeOBr\, as a function of time delay measured with the radiation electric field $\mathbf{E}^\omega$ perpendicular to the \emph{c}-axis ($\mathbf{E}^\omega\perp c$) at 3.9~K and 295~K. Following the first pulse, one can observe more pulses with much lower magnitude due to the multiple interference at the sample surface. The resulting transmission after Fourier transformation of the time-domain signal is shown in Fig.~\ref{Fig:TDS}(b). The observed transmission and phase correspond to a dielectric constant of about 12 at lowest frequencies. The spectrum at 295~K shows a strong absorption band from 7 to 11~meV, with transmission below the detection limit of the spectrometer. A periodic modulation on the transmission spectra can be seen in the whole frequency range due to the multiple interference effect. This absorption band is ascribed to phonons corresponding to the polarization of $\mathbf{E}^\omega$ $\perp$ \emph{c}, in agreement with the phonon spectra reported in the isostructural compound \CuBiSeOCl.\cite{Miller12} In the spectrum at 3.9~K, the absorption bands are clearly discriminated from each other. Two absorption bands are evidently observed at about 7.2~meV ($\sim58$~cm$^{-1}$) and 10.3~meV ($\sim83$~cm$^{-1}$). These values are slightly smaller than the eigenfrequencies of 68.3 and 89.0~cm$^{-1}$, respectively, where the corresponding phonons are observed in \CuBiSeOCl\, at 7~K.\cite{Miller12} This is consistent with the ratio $\sqrt{M_{CBSOC}}:\sqrt{M_{CBSOB}}=0.97$, where $M_{CBSOC}$ and $M_{CBSOB}$ are the molecular mass of Cu$_{3}$Bi(SeO$_3$)$_2$O$_2$Cl and Cu$_{3}$Bi(SeO$_3$)$_2$O$_2$Br, respectively. There are two other lower-lying phonons observed at 36.3 and 52.8~cm$^{-1}$ in \CuBiSeOCl.\cite{Miller12} The corresponding phonons cannot be clearly identified in the transmission spectrum of \CuBiSeOBr~due to multiple interference effects.

%Figure 2
\begin{figure}[t]
\centering
\includegraphics[width=88mm,clip]{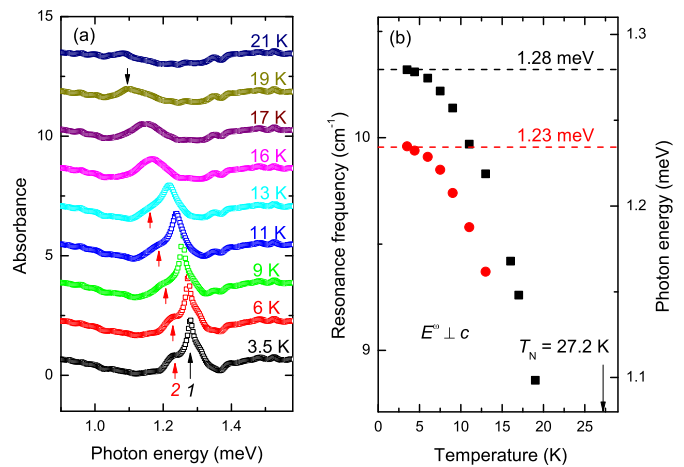}
\vspace{2mm} \caption[]{\label{Fig:AFMR_TDS} (Color
online) (a) Absorbance versus frequency measured at various temperatures below $T_N$ for $\mathbf{E}^\omega$ $\perp$ \emph{c}. The downward arrow marks a resonance mode observed at 19~K. The upward arrows mark two distinct modes observed at 3.5~K, which are named as mode 1 and mode 2, respectively. The curves are shifted with respect to the 3.5~K curve by a constant value for clarity. (b) Dependence of resonance frequencies on temperature, which reach the energies of 1.23 and 1.28~meV at 3.5~K. }
\end{figure}

In the spectrum at 3.9~K in Fig.~\ref{Fig:TDS}(b), a distinct feature can be observed at about 1.3~meV. Absorbance (Ab) is calculated from transmittance (Tr) via the relation $Ab=-\log_{10}Tr$. Figure~\ref{Fig:AFMR_TDS}(a) shows the absorbance versus frequency measured at various temperatures. An asymmetric peak (mode 1, black arrow) can be observed below 20~K in the antiferromagnetic phase. The peak shifts to higher frequency with decreasing temperature and two modes can be identified. At 3.5~K, the two modes are marked by arrows. The higher-lying mode 1 is sharper than the lower-lying mode 2. Above 15~K, mode 2 cannot be resolved anymore. The eigenfrequencies of both modes are plotted in Fig.~\ref{Fig:AFMR_TDS}(b) as a function of temperature. Since the modes are observed below the magnetic ordering temperature, the modes are assigned to magnetic excitations of the system. The hardening of the modes with decreasing temperature is reminiscent of the temperature dependence of a sublattice magnetization, and typical for magnetic resonance modes. The resonance frequencies of mode 1 and mode 2 at 3.5~K correspond to the energies of 1.28 and 1.23~meV, respectively.

\subsection{Frequency-domain THz spectra with applied magnetic field}
%Figure 3
\begin{figure*}[t]
\centering
\includegraphics[width=140mm,clip]{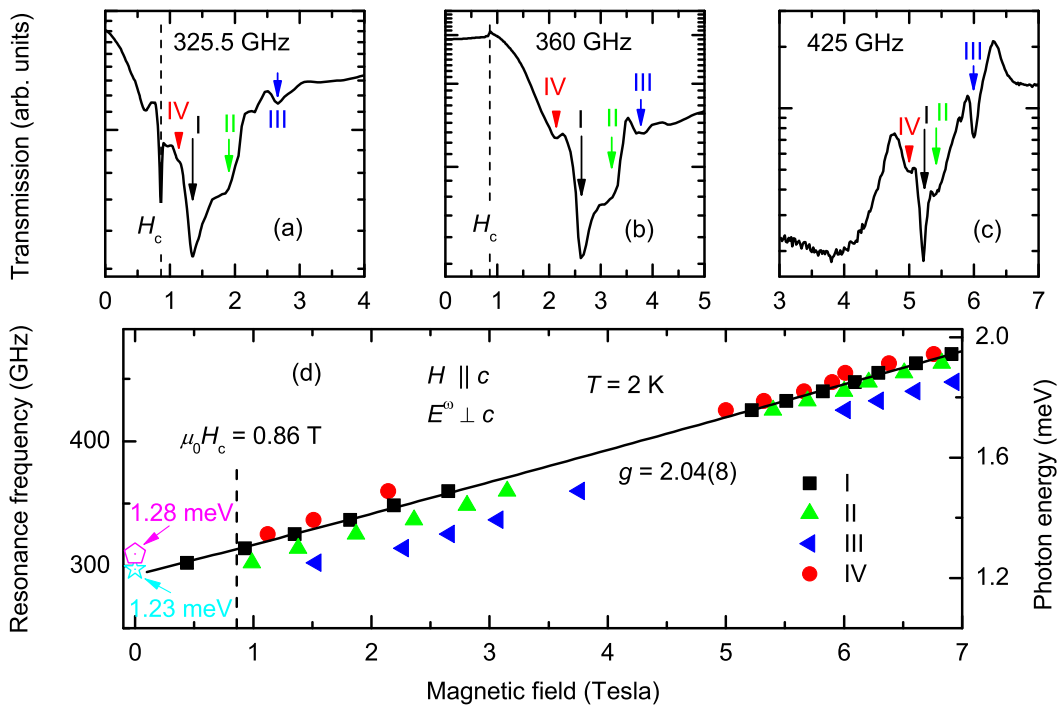}
\vspace{2mm} \caption[]{\label{Fig:FHDiag} (Color
online) (a)-(c) Transmission electron spin resonance spectra measured at 2~K with frequencies 325.5, 360, and 425~GHz, respectively. The external magnetic field $\mathbf{H}$ is applied along the \emph{c}-axis and the radiation electric field $\mathbf{E^\omega}$ is perpendicular to the \emph{c}-axis (Faraday configuration). The critical field is marked as $\mu_0H_c=0.86$~T. (d) Resonance frequencies versus resonance magnetic fields determined from various spectra.
The solid line is a fit to Eq.~(\ref{Eq:FerroMR_shape}) which determines a \emph{g}-factor of 2.04(8) (see text). The two zero-field modes from time-domain spectroscopy are shown by open symbols.}
\end{figure*}

Figures~\ref{Fig:FHDiag}(a)-(c) show the transmission spectra measured in a Faraday configuration ($\vec{k}$ $\|$ \textbf{H} $\|$ \emph{c}-axis) with various radiation frequencies at 2~K. The magnetic field $\mu_0\mathbf{H}$ up to 7 T is applied along the \emph{c}-axis. The radiation electric field $\mathbf{E^\omega}$ is perpendicular to the \emph{c}-axis ($\mathbf{E^\omega}\perp c$). An anomaly can be clearly seen in Fig.~\ref{Fig:FHDiag}(a) and (b) at the magnetic field of $\mu_0H_c=0.86$~T as indicated by the vertical dashed lines. This is consistent with the reported critical magnetic field of 0.8~T, above which the inter-layer ordering changes from antiferromagnetic to ferromagnetic.\cite{Pregelj12}

Above $H_c$, four resonance lines are observed in the transmission spectra at several frequencies. As marked by the arrows, one can see the strongest resonance mode at 5.21~T (mode I), and three other modes at 5.00, 5.40, and 6.01~T (mode IV, II, and III, respectively) from the transmission spectrum corresponding to 425~GHz radiation [Fig.~\ref{Fig:FHDiag}(c)]. By comparing the spectra of different frequencies [Fig.~\ref{Fig:FHDiag}(a)-(c)], we found that the resonance peaks shift to lower fields and become broader with decreasing frequency.  The resonance frequencies versus resonance fields are plotted in Fig.~\ref{Fig:FHDiag}(d). It clearly shows that all the four modes shift to higher frequency with higher magnetic field.

Neutron diffraction experiments have revealed that the magnetic moments of neighboring \emph{ab}-layers order ferromagnetically along the $c$ axis (easy axis) when $H>H_c$.\cite{Pregelj12} Above $H_c$ the total magnetic moment precesses about the direction of the static magnetic field, and energy is absorbed strongly from the incident radiation when its frequency is equal to the precessional frequency.\cite{Kittel7th} For an easy-axis compound, the magnetic anisotropy energy can be approximated by $K_1 \sin^2\theta$, when the quartic and higher-order terms of $\sin\theta$ are neglected, $K_1$ is the anisotropy constant, $\theta$ is the angle between the magnetization and the \emph{c} axis (easy-axis). In the ferromagnetic phase, the resonance frequency depends strongly on the sample shape due to demagnetization effect. When the magnetic field $H$ is applied along the easy axis, the dependence of resonance frequency on the magnetic field is given by\cite{Kittel7th,Gurevich96}
\begin{align}\label{Eq:FerroMR_shape}
\omega &=\frac{g\mu_B}{\hbar}\sqrt{(\mu_0H)^2+A\mu_0H+B^2},\\
\nonumber A&\equiv(N_x+N_y-2N_z)\mu_0M+2\mu_0H_A,\\
\nonumber B&\equiv\mu_0\sqrt{[(N_x-N_z)M+H_A][(N_y-N_z)M+H_A]},
\end{align}
where the magnetic anisotropy is considered, $H_A\equiv 2K_1/\mu_0\mu_{Cu}$ is the effective anisotropy field, $\mu_{Cu}$ is the magnetic moment per Cu ion, \emph{g} is the Land\'{e} factor, $\mu_B$ is the Bohr magneton, $\hbar$ is the Planck constant, $\mu_0$ is the vacuum permeability, and $N_x$, $N_y$, and $N_z$ are the demagnetization factors along the \emph{a}, \emph{b}, and \emph{c} axes, respectively. The magnetization $M$ in Eq.~(\ref{Eq:FerroMR_shape}) is, in general, a function of magnetic field. Above 0.86~T for $H \| c$, the magnetization reaches a saturation value corresponding to 0.9$\mu_B$ per Cu ion.\cite{Pregelj12} Therefore, we can assume that $M$ is constant when we fit the data above $H_c$.

We try to use Eq.~(\ref{Eq:FerroMR_shape}) to describe the field dependence of the strongest mode (mode I), which is shown by the solid squares in Fig.~\ref{Fig:FHDiag}(d). The fit to Eq.~(\ref{Eq:FerroMR_shape}) is illustrated by the solid line in Fig.~\ref{Fig:FHDiag}(d), which is in good agreement with the experimental data. The fit results in a \emph{g}-factor of 2.04(8), $A=17.1(21)$~T, and $B=10.3(4)$~T. This \emph{g}-factor is consistent with typical values for Cu$^{2+}$ ions.\cite{Deisenhofer06,Wang11b,Abragam1970} A consistent \emph{g}-value of 2.16 has been reported in the isostructural compound \CuBiSeOCl.\cite{Miller12}  The fitted \emph{B} value corresponds to the zero-field energy of 1.22(5)~meV. This is slightly smaller than 1.28~meV, where the dominant mode is observed in the time-domain spectra (mode 1 in Fig.~\ref{Fig:AFMR_TDS}). For the other three modes with much smaller intensity (Fig.~\ref{Fig:FHDiag}), the corresponding absorptions cannot be distinguished in the time-domain spectra.

%It is very likely that the mode shown by the upward solid triangles in Fig.~\ref{Fig:FHDiag}(d) corresponds to the 1.23~meV mode in Fig.~\ref{Fig:AFMR_TDS}, because this mode has a relatively .

The demagnetization factors $N_x=N_y=0$ and $N_z=1$ for an infinitely thin plate were used to estimate the anisotropy. Analyzing the experimental B value results in $K_1=0.27(1)$~meV, while the analysis of A results in $K_1=0.23(3)$~meV. Both values are consistent with the range $0.18<K_1<0.39$~meV estimated from the magnetization measurements via the relation $K_1=\mu_0\mu_{Cu}H_s/2$ due to the linear dependence of magnetization on magnetic field,\cite{Buschow03} where the saturation field $\mu_0H_s$ has been determined as 7~T for $\mathbf{H} \| b$ and 15~T for $\mathbf{H} \| a$.\cite{Pregelj12} It should be noted that deviations of the measured specimens' shapes from an infinite thin plate are additional sources of uncertainty for the estimated parameters.

The magnetic resonance modes are intimately dependent on the microscopic magnetic structure, including the exact spin configuration and exchange interactions. According to the partially solved spin configuration,\cite{Pregelj12} the spins on different sublattices should have different exchange anisotropies. The difference can be small and results in additional resonance modes with close energies, as we have observed from the modes I - IV. We note that the surface anisotropy of a ferromagnetic compound can also lead to additional resonance modes with small difference in resonance energies.\cite{Kittel7th}

%This roots in the variety of exchange paths mediated by the lone-pair ions, which leads to the geometric frustration in the pseudo-Kagome lattice.

In the isostructural compound \CuBiSeOCl, another magnetic resonance mode was observed at much higher frequency of about 1~THz $\sim$ 33cm$^{-1}$ with the same alignment of magnetic field, i.e., \textbf{H} $\|$ \emph{c}.\cite{Miller12} Besides being isostructural, \CuBiSeOBr\, has similar magnetic properties as \CuBiSeOCl, such as magnetic-ordering temperature, magnetic anisotropy, and a field-induced metamagnetic transition.\cite{Millet01,Pregelj12,Miller12} Thus, it is natural to expect that the resonance mode around 1~THz can be also observed in \CuBiSeOBr, but it is probably hidden by interference effect and the influence of the low-lying phonons in this range. Hence, there might be at least five magnetic resonance modes present in this system. All of these modes shift to higher frequencies with higher magnetic field,\cite{Miller12} and five or more sublattices will have to be considered to describe the complete magnetic resonance spectra.

\section{Conclusion}

In summary, the temperature-dependent time-domain THz spectra of single-crystal \CuBiSeOBr\, have been measured with the radiation propagation vector perpendicular to the sample \emph{ab}-plane. Two magnetic resonance modes have been resolved below the magnetic ordering temperature, where a paramagnetic to antiferromagnetic transition occurs. The two modes are hardening on decreasing temperature and reach energies of 1.28 and 1.23~meV at 3.5~K. Multi-frequency electron spin resonance transmission spectra have been measured in the Faraday configuration with the magnetic field parallel to the sample \emph{c}-axis. Another two magnetic resonance modes are observed in these spectra. The dependence of the resonance modes on magnetic field are studied at 2~K, which reveals a monotonic increase of the resonance frequency with increasing magnetic fields. The demagnetization effect and exchange anisotropy have been considered to describe the field dependence of the dominant magnetic resonance mode, resulting in a \emph{g}-factor of 2.04(8) and a zero-field resonance frequency of 1.22(5)~meV. The magnetic anisotropy is estimated to be about 0.25~meV. In addition, a metamagnetic transition from antiferromagnetic to ferrimagnetic phase has been confirmed to occur at 0.86~T. These results are in agreement with those from neutron diffraction and magnetic susceptibility measurements.

\begin{acknowledgments}
We thank A. G\"{u}nther for the fruitful discussions. We acknowledge partial support by the Deutsche Forschungsgemeinschaft via TRR 80
(Augsburg-Munich), FOR 960 (Quantum Phase
Transitions) and project DE 1762/2-1.
\end{acknowledgments}

\end{document}